\documentclass[floatfix,prd,showpacs,twocolumn]{revtex4}

\usepackage{amssymb}
\usepackage{amsmath}
\usepackage[dvips]{graphicx}
\usepackage[dvips]{color}

\begin{document}

\title{The relativistic glider}

\author{Eduardo Gu\'eron}
\email{eduardo.gueron@ufabc.edu.br}
\affiliation{Centro de Matem\'atica, Computa\c c\~ao e Cogni\c c\~ao, Universidade
  Federal do ABC, 09210-170, Santo Andr\'e, SP, Brazil}

\author{Ricardo A. Mosna}
\email{mosna@ime.unicamp.br}
\affiliation{ Instituto de Matem\'atica, Estat\'\i stica e Computa\c{c}\~ao Cient\'\i fica,
Universidade Estadual de Campinas, C.P. 6065, 13083-859, Campinas, SP, Brazil.}

\begin{abstract}
We present a purely relativistic effect according to which
{\em asymmetric} oscillations of a quasi-rigid body slow down
or accelerate its fall in a gravitational background.
\end{abstract}

\pacs{04.20.-q, 04.25.-g, 45.40.-f }
\maketitle

The motion of extended bodies in General Relativity has been a
source of unexpected phenomena up to the present. Although the
problem in its full generality is notoriously
complicated~\cite{dixon}, approximated analysis of quasi-rigid
bodies already discloses interesting effects. A remarkable example
and quite surprising result was recently presented by Wisdom, who
demonstrated that cyclic changes in the shape of a quasi-rigid body
may lead to net translation in curved spacetimes~\cite{W}. Since
this takes place in the absence of external forces, one can portray
the body as swimming in spacetime. Later, this effect was shown to
dominate over the so-called ``swinging'' (Newtonian) effect, whose
formulae are well known, for high enough oscillation
frequencies~\cite{GMM}. The latter is a direct expression of the
fact that expanding and contracting a body in a nonuniform
gravitational field gives rise to a net work on it (due to tidal
forces). However, this effect is increasingly suppressed for higher
frequencies, since the gravitational potential felt by the body is
then nearly the same through its complete cycle (see Eq.~5 of \cite{GMM}).

Here we present a purely relativistic effect which (like the
swimming and unlike the swinging effect) is not suppressed for high
frequencies, but (like the swinging and unlike the swimming effect)
is not related to a gravitational geometric phase.
It comes about by letting a quasi-rigid body perform {\em asymmetric}
oscillations in a gravitational background. We show that, as a result,
the body slows down or accelerates its fall depending on how asymmetric
its internal oscillations are.

We propose a simple model to investigate this effect:
two test particles of  mass $m$ connected by a massless strut whose
length varies periodically, i.e., an oscillating dumbbell.
The strut is mathematically implemented as a constraint between the
particles. Physically, and following Wisdom's idea (see last section
of \cite{W}), the active strut should be constantly and locally
monitored in order to guarantee that the constraints are maintained.

A test particle under the influence of a spherically symmetric gravitational
field moves, in the context of General Relativity, in Schwarzschild
spacetime. Its equations of motion follow from the Lagrangian
\begin{equation}
{\mathcal L_p}= - m c
    \sqrt { \left( 1-\frac{2GM}{c^2 r_p}\right)c^2\left(\frac{dt}{ds}\right)^2 -
    \frac{1}{1-\frac{2GM}{c^2 r_p}}
    \left(\frac{dr_p}{ds}\right)^2},
\label{lag1}
\end{equation}
where $m$ is the mass of the test particle and $s$ is an evolution parameter
(we assume that the body falls vertically towards the gravitational source).
For the dumbbell, we choose the coordinate time $t$ as the evolution
parameter and take the relevant Lagrangian as
\begin{widetext}
\begin{equation}
{\mathcal L_d}=
-mc \sqrt {\left(\!\! 1-\frac{2GM}{c^2 r_1}\!\!\right)c^2 -
  \frac{1}{1-\frac{2GM}{c^2 r_1}} \left(\!\!\frac{dr_1}{dt}\!\!\right)^2 }
-  \left.  mc \sqrt{\left(\!\!1-\frac{2GM}{c^2 r_2}\!\!\right)c^2 - \frac{1}{1-\frac{2GM}{c^2  r_2}}
  \left(\!\!\frac{dr_2}{dt}\!\!\right)^2 }\,\right|_{r_2(t)=r_1(t)+l(t)} \hspace{-2cm},
\label{lag2}
\end{equation}
\end{widetext}
where $l(t)$ is the constraint as seen by the observer at infinity.
In the Newtonian limit, $l(t)$ represents the length of the strut
connecting the particles at time $t$. This implementation of the
constraint clearly embodies considerable simplification
(for instance, it fails to be covariant)~\cite{fn:spgeodesic}.
However, it is important to note that, by choosing a periodic $l(t)$
such that $l(0)=l(T)=0$, all observers will agree that the worldlines
of both particles meet at the end of each period.
In the following, we numerically solve the equations of motion coming from~(\ref{lag2})
for a family of constraint functions $l_\alpha(t)$ whose members
always meet this condition. Of course, different observers will disagree
on the symmetry and shape of a given function $l(t)$ as measured by them.

With these considerations in mind, we compute the fall of the
oscillating dumbbell and compare it with that of the single particle
(the particular case when $l(t)\equiv 0$), both with the same initial conditions.
Given that the cyclic changes considered here affect a single
degree of freedom, we can anticipate that the swimming effect will
not arise in this case~\cite{fn:geomphase}.
Therefore the sought after relativistic effect will be isolated
once we work in a regime for which (Newtonian) tidal effects are negligible.

\begin{figure}[h]
\begin{center}
\includegraphics[width=\linewidth]{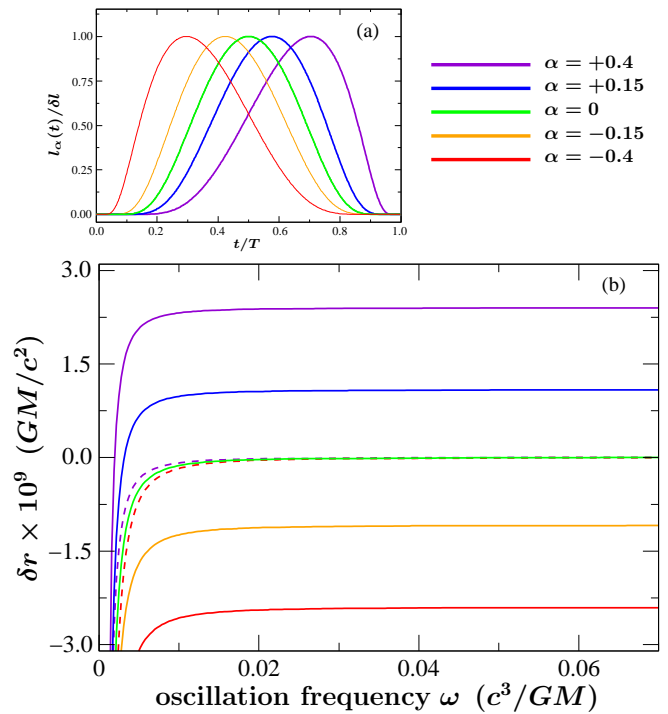}
\caption{(a) Family of constraint functions $l_\alpha(t)$~\cite{fn:shapes}.
(b) Net displacement $\delta r = r_1(T)-r_p(T)$ of the dumbbell, relative to the falling
point mass, after one period.
The initial conditions are set by $\dot{r_p}(0)=\dot{r}_1(0)=0$, $r_p(0)=r_1(0)=120$
and $\delta l=5\times 10^{-3}$ (in units of $GM/c^2$).
The relativistic (Newtonian) case is represented by solid (dashed) lines.
The plot shows that the Newtonian $\delta r$ is always negative
and that it is negligible for $\omega\gtrsim 0.03 \ c^3/GM$, irrespective of $\alpha$.
In contrast, we see that the relativistic $\delta r$ crucially depends on
the asymmetry parameter $\alpha$, and does attain positive values
(for positive $\alpha$). Causality problems (due to large velocities) only
start to appear at much higher frequencies, not shown above.
}
\label{fig:ls}
\end{center}
\end{figure}

We numerically solved the Euler-Lagrange equations associated with~(\ref{lag2})
for the family $\{l_{\alpha}(t)\}$ portrayed in Fig.~\ref{fig:ls}a.
The asymmetry parameter~$\alpha$, ranging from $-1$ to $1$, quantifies how much
$l_\alpha(t)$ fails to be symmetric about its peak:
the function corresponding to $\alpha=0$ has its peak at
$t=T/2$, while a generic $l_\alpha(t)$ attains its maximum
at $t=\tfrac{1+\alpha}{2}\, T$~\cite{fn:shapes}.
For each value of $\alpha$, the equations of motion of the oscillating
dumbbell and of the point mass were integrated, with $r_p(0)=r_1(0)\equiv R$ and $\dot{r_p}(0)=\dot{r}_1(0)=0$.
Their separation $\delta r = r_1(T)-r_p(T)$ after one period then yields the net displacement of the dumbbell.
Representative results, for $R=120$ and $\delta l=5\times 10^{-3}$, in units of $GM/c^2$,
are shown in Fig.~\ref{fig:ls}b, where $\delta r \times 10^9$
was plotted against the dumbbell oscillation frequency $\omega=1/T$.
The relativistic case, corresponding to (\ref{lag1}) and (\ref{lag2}),
is represented by solid lines, while its Newtonian counterpart is depicted
by dashed lines.

Fig.~\ref{fig:ls}b shows that the Newtonian $\delta r$ is always negative and
that it becomes negligible when the frequency is large enough
(as expected from our earlier discussion).
It is important to note that this happens
{\em irrespective of the value of the asymmetry parameter $\alpha$}.
This is in marked contrast to the relativistic case.
Indeed, while the relativistic curve for the symmetric
case ($\alpha=0$) is surrounded by Newtonian curves, those corresponding to
positive values of $\alpha$ lead to a {\em positive} net displacement,
a purely relativistic effect.
Therefore, a relativistic robot who cannot swim may still
soften its fall by repeatedly stretching and shrinking its
body in an asymmetric fashion.
Note that the relativistic $\delta r$ does
not become negligible for large values of $\omega$, and that
its sign and magnitude in this regime are determined by $\alpha$.

It is worth noting that the beginning of the plateaus in Fig.~\ref{fig:ls}b
occurs in a frequency domain which is fairly independent on the
oscillation amplitude $\delta l$~\cite{fn:indep}.
By keeping $\delta l$ small one can thus obtain the effect in hand for reasonably
small velocities.
It is interesting to note that, by Taylor expanding the Lagrangian~(\ref{lag2})
in $\dot{r}_1$ and $\dot{r}_2$ while keeping the remaining (position-dependent) terms
unperturbed, one needs to retain terms at least up to fourth order in $v/c$ to capture
the effect under consideration.
This should be compared to Wisdom's swimming effect, which already takes
place when the relevant Lagrangian is expanded up to second order in $v/c$.
The effect presented here thus arises from an intricate coupling
of the gravitational field to the velocities of the particles.

\begin{figure}[h]
\begin{center}
\includegraphics[width=0.8\linewidth]{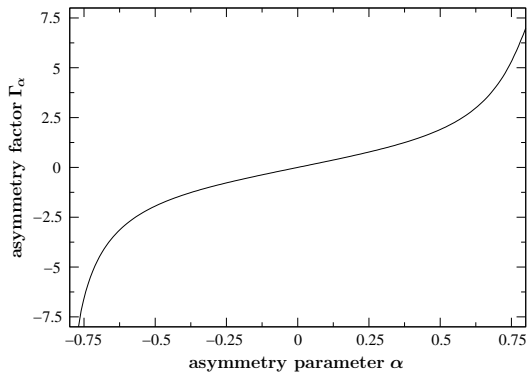}
\caption{Adimensional scale factor $\Gamma_{\!\alpha}$ as a function
  of the asymmetry parameter $\alpha$.}
\label{fig:gamma}
\end{center}
\end{figure}

From some numerical analysis, we estimate that the net displacement
after one period, in the large frequency regime (corresponding to the
plateaus in Fig.~\ref{fig:ls}), is given by
\begin{equation}
\delta r \approx \Gamma_{\!\alpha} \frac{\delta l^2}{R^2} \; \frac{GM}{c^2},
\label{deltar}
\end{equation}
where $\delta l$ is the amplitude of $l_\alpha(t)$ and $\Gamma_{\!\alpha}$ is a
dimensionless factor associated with the asymmetry parameter $\alpha$,
see Fig.~\ref{fig:gamma}.  This expression was obtained from
numerical simulation employing a large range of values for $R$ and
$\delta l$, going from $50$ to $10^7$ and $10^{-7} R$ to $10^{-2} R$,
respectively.
Equation~\ref{deltar} shows that the effect under consideration can be
orders of magnitude higher than the swimming effect,
which is proportional to $\delta l^2/R^3$~\cite{fn:intensity}.
This result is somewhat surprising since, as discussed above, the
swimming effect
is already present when the Lagrangian is expanded up to second order
in the velocities, while the effect presented here needs at least
a fourth-order expansion in $v/c$.

It is instructive to calculate the net displacement $\delta r$ in terms of the
coordinate distance $D$ traveled by the falling point mass during the
period $T$.
Using the rough estimate $D \sim g T^2$, with $g = GM/R^2$, we immediately
see that
\begin{equation}
\delta r/D \sim (\delta l \,\omega/c)^2
\end{equation}
 in this regime. In this way, if the dumbbell is engineered so that
$\delta l \, \omega \sim 0.01c$, we get $\delta r/D \sim 10^{-4}$.
Therefore, such result may be within reach of experimental verification.
This would open the door to new tests of General Relativity by means of
local experiments~\cite{fn:earth}.

We close by noting that, if $\delta r/D$ were to reach $1$,
the dumbbell would be able to float in space. However, this is apparently
not possible, since numerical simulations indicate that the value attained
by $\delta r/D$ is typically limited to $15\%$.
Therefore, it seems that, although this mechanism cannot be employed to fly,
it might be conceivably used as a relativistic paraglider to slightly
soften the landing on an atmosphereless planet.

\acknowledgments

The authors thank G. E. A. Matsas, S. R. Oliveira and Alberto Saa for fruitful
discussions. EG is grateful to P. S. Letelier for the technical support.
RAM acknowledges financial support from FAPESP.


\begin{thebibliography}{99}
\bibitem{fn:spgeodesic}
A related (and also simplified) quasi-rigidity condition could be defined
by imposing that the geodesic length of constant $t$ slices connecting the worldlines
of the particles be given by the constraint function $l(t)$.
The obtained results are essentially the same for either of these choices.
\bibitem{fn:shapes}
The functional form of the constraint functions in Fig.~\ref{fig:ls}a
is given by
$l_\alpha(t) = \delta l \exp\left[ \frac{(1+\alpha-2\omega t)^2}{(1+\alpha^2)\omega t(-1+\omega t)} \right]$,
where $\omega=1/T$ and the overall normalization was chosen so
that the amplitude of $l_\alpha(t)$ is $\delta l$.
Other families $\{l_\alpha(t)\}$ with diverse shapes were also tested,
with no appreciable changes in the results.
\bibitem{fn:indep}
Numerical simulations indicate that a variation in $\delta l$ of five
orders of magnitude leads to a change of less than $1\%$ in the characteristic
frequency for which the corresponding plateau begins. 
\bibitem{fn:earth}
We note that, if $M$ is the Earth's mass, then $R \sim 10^9 \, GM/c^2$ at Earth's surface.
Although we could numerically confirm Eq.~(\ref{deltar}) only for
$50 \, GM/c^2 \lesssim R \lesssim 10^7 \, GM/c^2$
(due to limitations of our integration routine), there is no indication that it
would break down at such scale.
\end{thebibliography}
\end{document}